\documentclass[rmp,nofootinbib,twocolumn]{revtex4}
\usepackage{graphicx}
\ifnum\lefthyphenmin<2\lefthyphenmin=2\fi
\ifnum\righthyphenmin<2\righthyphenmin=2\fi
\bibpunct[, ]{[}{]}{,}{a}{,}{,}

\newcommand{\Celsius}{^\circ {\tt C}}

\newcommand{\kcal}{{\tt kcal/mole}}
\newcommand{\kT}{k_BT}
\newcommand{\betainv}{\beta^{-1}}
\newcommand{\lambdainv}{\lambda^{-1}}
\newcommand{\bp}{{\tt bp}}
\newcommand{\AT}{{\tt AT}}
\newcommand{\GC}{{\tt GC}}

\newcommand{\eps}{\varepsilon}

\newcommand{\Tau}{{\cal T}}
\newcommand{\init}{\gamma_1}
\newcommand{\Gd}{\Tau_{\rm d}}
\newcommand{\SingleBubbleZ}{Z}
\newcommand{\SingleBubbleF}{F}
\newcommand{\SingleBubbleE}{E}
\newcommand{\SingleBubbleS}{S}
\newcommand{\AveSingleBubbleZ}{\overline{\SingleBubbleZ}}
\newcommand{\AveSingleBubbleF}{\overline{\SingleBubbleF}}
\newcommand{\AnnealSingleBubbleF}{\widetilde{\SingleBubbleF}}
\newcommand{\AnnealSingleBubbleE}{\widetilde{\SingleBubbleE}}
\newcommand{\AnnealSingleBubbleS}{\widetilde{\SingleBubbleS}}
\newcommand{\AnnealSingleBubbleL}{\widetilde{L}}
\newcommand{\SingleBubbleH}{{H}}
\newcommand{\cL}{L}
\newcommand{\FixedLengthZ}{{\cal Z}}
\newcommand{\FixedLengthF}{{\cal F}}

\newcommand{\AveFixedLengthF}{\overline{\FixedLengthF}}
\newcommand{\AveFixedLengthZ}{\overline{\FixedLengthZ}}
\newcommand{\AnnealFixedLengthF}{\widetilde{\FixedLengthF}}

\newcommand{\AnnealFixedLengthS}{\widetilde{{\cal S}}}
\newcommand{\FixedLengthH}{{\cal H}}
\newcommand{\M}{\mathsf{M}}
\newcommand{\bM}{\mbox{\boldmath$\M$}}

\newcommand{\dG}{\delta G}
\newcommand{\DG}{\Delta G}

\newcommand{\ValleyG}{\Delta\widehat{G}}
\newcommand{\ValleyL}{\widehat{L}}
\newcommand{\MinE}{E^*}
\newcommand{\MinL}{L^*}
\newcommand{\MinH}{H^*}
\newcommand{\AveMinE}{\overline{\MinE}}
\newcommand{\AvedG}{\overline{\dG}}
\newcommand{\AvedGG}{{\rm var}(\dG)}

\newcommand{\Nx}{N_\times}
\newcommand{\AnnealNx}{\widetilde{N}_\times}
\newcommand{\MinNx}{N^*_\times}
\newcommand{\W}{\widetilde{W}}

\newcommand{\dt}{\Delta t}

\begin{document}

\title{Localization of Denaturation Bubbles in Random
DNA Sequences }

\author{Terence Hwa$^{1}$}
\author{Enzo Marinari$^{2}$}
\author{Kim Sneppen$^{3}$}
\author{Lei-han Tang$^{4}$}
\affiliation{
$^{1}$Department of Physics, 
University of California at San Diego,
La Jolla, CA 92093-0319\\
$^{2}$Dipartimento di Fisica, SMC and UdR1 of INFM, 
Universit\`a di Roma {\em La Sapienza}, P. A. Moro 2, 
00185 Roma, Italy\\
$^{3}$Niels Bohr Institute, Blegdamsvej 17, 2100 Copenhagen, Denmark \\
$^{4}$Department of Physics, Hong Kong Baptist University, Kowloon Tong,
Hong Kong
}

\date{\today}

\begin{abstract}
We study the thermodynamic and dynamic 
behaviors of twist-induced denaturation bubbles in a long, stretched
random sequence of DNA. The small bubbles associated with weak twist
are delocalized. Above a threshold torque, 
the bubbles of several tens of bases or larger
become preferentially localized to \AT-rich segments.
In the localized regime, the bubbles exhibit ``aging'' and
move around sub-diffusively with continuously varying 
dynamic exponents. These properties are derived
using results of large-deviation theory together with 
scaling arguments, and are verified by Monte-Carlo simulations.
\vspace{-1in}
\end{abstract}

\maketitle

%\begin{multicols}{2}

\section{Introduction}
\vspace{-10pt}
Localized opening of double-stranded DNA is essential in a number
of cellular processes such as the initiations of gene transcription
and DNA replication~\cite{Alberts}. While thermal denaturation is highly 
unlikely under physiological conditions, {\it in vitro} experiments 
show that local denaturation can be readily induced by 
{\em under-winding} the DNA double helix by an amount that
is physiologically reasonable~\cite{Kowalski,Cozzarelli,Bensimon}. 
The basic physical effect is simple: An under-wound double helix 
suffers a reduction in the binding free energy~\cite{Marko,Cocco,Strick00}. 
Local openings of the double helix (referred to as ``denaturation bubbles'') 
relieve the twist experienced by the remainder
of the double helix and is thus energetically favorable. 
The denaturation bubbles may be recruited to a specific location of the genome 
by a designed (e.g., \AT-rich) sequence, 
since \AT\ pairs bind more weakly than \GC\ pairs~\cite{Benham}. 
On the other hand, entropic effect which 
favors bubble {\em delocalization}  is  non-negligible
for long sequences. Also significant is the kinetic trapping 
of the bubbles due to statistical agglomeration of \AT-rich segments
in long heterogenous sequences.

To gain some quantitative understanding on the competing effects
of entropy and sequence heterogeneity, 
we characterize in this study the  thermodynamic and dynamic
properties of denaturation bubbles in a long, stretched {\em random}
DNA sequence with no special sequence design. 
%We investigate the conditions under which the 
%denaturation bubbles are localized (i.e., trapped) by the random sequence,
%and characterize their sizes and dynamics in the localized regime.
Previously, there have been a number of experimental and theoretical
studies~\cite{Cule,tang01,Heslot,Lubensky} on the effect of sequence 
heterogeneity on
DNA melting and unzipping transitions. Our study is along this general 
direction. The specific behaviors exhibited by the denaturation bubbles
are rather complex and are typical of those observed in
systems dominated by quenched disorders~\cite{glass}:
The bubbles are localized upon increase of the
applied torque beyond a certain threshold. In the localized regime,
their dynamics exhibits ``aging''~\cite{aging1,aging}
and is sub-diffusive with continuously varying exponents.

Interestingly, twist-induced denaturation presents
a rare physical example of the celebrated
random-energy model of disordered system~\cite{rem}.
Consequently,  detailed analysis of both the
thermodynamis and dynamic properties can be made
by applying the well-developed theory of disordered systems~\cite{glass},
together with exact results from large-deviation theory
familiar in the related sequence alignment problem~\cite{KarlinAltschul,YuHwa}.
We will draw upon detailed experimental knowledge of thermal 
denaturation~\cite{SantaLucia,SantaLucia1,Blake} throughout the analysis,
and make our results quantitative whenever possible.

\section{Thermodynamics}

Let us consider the application of a torque which under-winds a long,
stretched\footnote{A modest stretching force is needed
to prevent the applied torque from being absorbed by super-coiling; see 
e.g., Ref.~\cite{Cocco}.} 
piece of double-stranded DNA. We are interested in the regime where
the applied torque $\Tau$ is {\em below} the threshold $\Gd$
for bulk denaturation, but sufficiently strong so that 
denaturation bubbles do appear in the system.  
%We wish to characterize the typical
%sizes of the bubbles, their typical separation distances and
%dynamics, for random nucleotide sequences at different
%$\Tau$'s.  We will take advantage of 
Due to the highly cooperative nature of the denaturation process, 
the typical distance $N_\times$ between the large bubbles is large,
in which case treating the bubbles as a dilute gas of particles is 
appropriate. Our strategy will be to first characterize
analytically the thermodynamic behavior of a {\em single}
bubble, and then use this knowledge to determine
the length scale $N_\times$ and the many-bubble states for $N \gg
N_\times$. We will find that $N_\times > O(10^3)\, \bp$ as long as we
are not very close to the threshold $\Gd$, so that the dilute gas
approximation is reasonable  for a large range of parameters.

\subsection{The Single-bubble Model}

Consider a denaturation bubble confined in a DNA double helix
between two complementary DNA strands of $N$ bases each.
The double-strand is denoted by the base sequence 
$b_1b_2\ldots b_N$ [with $b_k \in \{{\tt A, C, G, T}\}$]
of one of the strands, ordered from 
the 5'- to 3'- end. %as shown in Fig.~\ref{F-BUBBLE}.
%
%\begin{figure}[t]
%\includegraphics[width=2.0in]{loop}
%  \caption{A bubble of length $L=4$ open base pairs
% in a double-stranded DNA
%    of $N=14$ base pairs. The first and last open pairs of the
%bubble are denoted by the index $m$ and $n=m+L-1$ respectively.
%    \protect\label{F-BUBBLE}
%  }
%\end{figure}
%
To simplify the notation, we assume that the two ends of the helix is sealed,
%imagine that there are two fictitious bound pairs at positions $0$ and $N+1$
so that the bubble is always contained in the segment $b_1\ldots b_N$. 
%[We are interested in the limit of long sequences ($N\gg 1$)
%for which the precise treatment of the boundaries is not important.]
Let the index of the first and last open pairs
of the open bubble be $m$ and $n$, with $ 1 \le m \le n \le N$.
We denote the total free energy of the bubble (defined with respect
to the helical state) by $\DG_L(m)$, where  $L \equiv n-m+1$ is the
number of open bases and referred to as the bubble length. Then 
the partition function of the single-bubble system is given by
\begin{equation}
  \SingleBubbleZ(N) = \sum_{L=1}^{N} \sum_{m=1}^{N-L+1} 
e^{-\beta \DG_L(m)}\;,
  \label{SingleBubbleZ-def}
\end{equation}
where $\betainv \equiv \kT\simeq 0.62~\kcal$ at $37 \Celsius$. 
%Thus, once the bubble energy $\DG_L(m)$ is specified, 
%all thermodynamic quantities
%can be computed from (\ref{SingleBubbleZ-def}).

%The bubble energy  clearly depends on the
%composition of the subsequence $b_m b_{m+1} ... b_n$.  
In the absence of the external torque, the bubble energy $\DG_L(m)$
has two components.
First, there is the loss of {\em stacking energy} $\dG_{b,b'}$
between two successive bases $b$ and $b'$. These stacking energies
are in the range $0.5 \sim 2.5\,\kT$'s at $37\Celsius$,
with the \AT\ stacks weaker than the \GC\ stacks. Their values have
been measured carefully~\cite{SantaLucia,SantaLucia1,Blake}.
Second, assuming that there is no secondary pairing between 
bases in the bubble so that the open configuration can be regarded
as a polymer loop, there is a well-known
polymeric {\em loop entropy} cost
\begin{equation}
\gamma_\cL = \init + \alpha \cdot \kT\; \ln \cL
\label{gamma}
\end{equation}
for a bubble of length $\cL$, with 
$\alpha \approx 1.8$~\cite{MEFisher} for a linearly extended
\footnote{The value of $\alpha$ may well be different for
{\em unstretched} DNA chain and hence relevant for the thermal
denaturation of DNA~\cite{Mukamel,Orland}. 
However, as we show below, essential features 
of denaturation do not hinge on the precise value of $\alpha$.} 
DNA chain. The bubble initiation cost $\init$ depends on
the base composition at opening and closing ends, ionic strength etc., 
and generally lies\footnote{The initiation cost for DNA bubbles
are extracted from the webserver: 
{\it http://www.bioinfo.rpi.edu/applications/mfold/} (M.\ Zuker, private
communication).
See also Ref.~\cite{Rouzina} for an alternative source.} 
in the range $3\sim 5\,\kT$.
For relevant bubble sizes of few tens of bases in length (see below),
the total entropic cost is $\gamma_\cL = 8 \sim 12\, \kT$.
This large cost justifies
the single bubble approximation (at least to the length scale
$N_\times \sim e^{\beta \gamma_\cL} \gtrsim 5\times 10^3\, \bp$), 
and contributes significantly to the sharpness of the observed
thermal denaturation transition~\cite{Blake}.

An applied negative torque $\Tau$ reduces the thermodynamic
stability of the helical state relative to the denatured one by
an amount equal to the work done to unwind the helix.
This effect is simply modeled here by a {\em linear} decrease in the 
stacking energy in the relevant parameter range~\cite{Cocco}, i.e.,
$\dG_{b,b'} \to \dG_{b,b'} - \theta_0\, \Tau$,
where $\theta_0=2\pi/10.35$ is the twist angle per base of the double helix.
Putting the above together, we have
\begin{equation}
\DG_L(m) = \gamma_\cL + \sum_{k=m}^{m+L} \dG_{b_{k-1},b_k} 
- \theta_0\, \Tau\cdot L
\label{E-def}
\end{equation}
as the single bubble energy, which can be computed 
once the DNA sequence $b_1...b_N$ is given.
Note that while Eq.~(\ref{E-def}) is formulated specifically
for twist-induced denaturation, the general form can be used to
describe a number of destabilizing effects, e.g.,  
due to changes in temperature, ionic concentration, etc.

\subsection{Sequence Heterogeneity}

As the torque $\Tau$ increases from zero towards the denaturation 
point $\Gd$,
denaturation bubbles appear in the double strand and grow in size. 
We wish to know whether the bubbles are free to 
diffuse along the double strand, or are they localized in the high
\AT\ regions of the DNA where binding is the weakest.
For simplicity, we will characterize the typical behavior of
an ensemble of random (i.e., independent and identically distributed) 
sequences described by the single-nucleotide frequencies $p_b$,
although our approach and qualitative findings are
also applicable to sequences with short-range correlations.  

For a {\em given} sequence of bases, 
the partition function $\SingleBubbleZ$ can of course be 
efficiently evaluated
numerically  (including all the multiple-bubble states)
by using available programs such as MELTSIM~\cite{Blake}.
All thermodynamic quantities can subsequently
be evaluated from the free energy 
$\SingleBubbleF = -\kT \ln \SingleBubbleZ$.
To obtain the {\em typical} behavior of the ensemble, we
ideally want to compute the ensemble average of the free energy,
$\AveSingleBubbleF \equiv -\kT \, \overline{\ln \SingleBubbleZ}$. 
[We use the over-line to denote average over the ensemble 
of random sequences, i.e., 
$\overline{X} \equiv \sum_{b_1,  ..., b_N} X_{b_1, ..., b_N} 
\prod_{k=1}^N p_{b_k}$; this is also known as the 
``disorder average''.] Computing $\AveSingleBubbleF$ 
numerically however
will require explicit generation of a large number of random
sequences and can be very time consuming for large $N$'s.
Fortunately we can apply a large body of knowledge accumulated
from the statistical mechanics of random systems~\cite{glass} and 
provide a detailed characterization of the typical behavior of our
system without the need of exhaustive simulation. 
To introduce notation and concepts in this approach,
we examine first the simplified problem of a single bubble
with a {\em fixed} length. 
%This will allow us to 
%introduce notation and concepts, 
%and develop an intuitive understanding of the system.
%It will then be straightforward to extend this approach to 
%bubbles without length restriction.\\

%                                                                 
\subsection{Bubble with Fixed Length}
Let us consider a bubble with a fixed length $\cL$ 
(with $1 \ll \cL \ll N$) embedded
in a long, random sequence $b_1...b_N$. The partition function reads
\begin{equation}
\FixedLengthZ_\cL = \sum_{m=1}^{N-\cL+1} \exp[-\beta \DG_\cL(m)],
\label{FixedLengthZ-def}
\end{equation}
where  the scripted variables refer to properties of 
the fixed length bubble. For a random sequence, the energies of
the different states labeled by $m$ are uncorrelated with each
other beyond the distance $\cL$. Such systems belong
\footnote{The correlation in $\DG$ between neighboring states is 
only a minor complication because it is short-ranged and can be transformed 
away by coarse graining.}
%that can be dealt with systematically in a block transformation.
%Specifically, we may group states into equal-sized blocks, each block 
%covers a sequence interval of size much bigger than $L$. 
%The overlap of states from adjacent blocks is seen to generate a boundary 
%term in the block free energy whose importance
%diminishes as the block size grows. Mapping to the pure REM behavior
%can be understood in this ``renormalization group'' sense.
%}
to the class of
``Random Energy Model'' (REM) and was solved 
exactly in the 80's by Derrida~\cite{rem} for a Gaussian distribution of 
$\DG$'s. Discrete distribution of $\DG$'s was studied in the 
closely-related system involving protein-DNA interaction~\cite{Gerland}.  
Below, we will briefly review the salient 
properties of REM using the present example.

The REM has a ``high-temperature'' phase where many (of order $N$)
bubble configurations contribute significantly to the partition sum, and a 
``low-temperature'' phase dominated by only one or a few lowest
energy states. It follows that, in the former case, the bubble is 
{\em delocalized} and can freely diffuse along the sequence, while
in the latter case, the bubble is {\em localized} to the
lowest energy position.
Transition between the two phases is driven by competition
between the energetic (variation in $\DG$) and entropic ($\ln N$)
effects. In the present problem, the magnitude
of terms in the partition sum (\ref{FixedLengthZ-def}) can be tuned
not only by varying the temperature,
but also by varying the bubble size $L$.
Hence, at a fixed $\beta$, whether a bubble is free or localized
depends both on the bubble size $L$ and the sequence length $N$.

An interesting property of the REM is that, in the ``high-temperature''
phase, $\FixedLengthZ_\cL/N$ tends to a {\rm finite limit} given by the
{\em annealed average} $\AveFixedLengthZ_\cL/N$ as $N\rightarrow\infty$, 
independent of the particular realization of the random sequence.
This allows us to replace the average free energy
$\AveFixedLengthF \equiv -\kT\, \overline{\ln \FixedLengthZ}$ 
by its {\em annealed approximation}
$\AnnealFixedLengthF \equiv - \kT \ln \AveFixedLengthZ$,
which is much easier to calculate. [We will use the
tilde to denote all quantities computed in the annealed approximation.]
Introducing a $4\times 4$ matrix  $\bM(\beta)$ with components 
$\M_{b,b'} = \sqrt{p_b p_{b'}}\exp[-\beta\; \dG_{b,b'}]$, and let
the largest eigenvalue of $\bM(\beta)$ be $\Lambda(\beta)$,
then the disorder average of terms in (\ref{FixedLengthZ-def})
can be written as 
\begin{equation}
%\overline{\exp\left[-\beta \sum_{k=m}^{m+\cL-1}\dG_{b_{k-1},b_k}\right]}
\overline{e^{-\beta \sum_{k=m}^{m+\cL-1}\dG_{b_{k-1},b_k}}}
= {\rm Tr}\ \bM^\cL(\beta) \stackrel{L\gg1}{=} \Lambda^\cL(\beta).
\label{M}
\end{equation}
It is convenient to introduce the quantity
\begin{equation}
f(\beta) \equiv  -\betainv \ln \Lambda(\beta), \label{f}
\end{equation}
with which we have (for $N\gg L$)
\begin{equation}
\AveFixedLengthZ_\cL =  \frac{N\, e^{-\beta\gamma_1}}{\cL^\alpha} \,
\left[e^{-\beta [f(\beta)- \theta_0\Tau]}\right]^{\cL}.
\label{AveFixedLengthZ}
\end{equation}
Hence, in the delocalized phase, 
$\AveFixedLengthF\simeq \AnnealFixedLengthF=
-\kT \ln N+L[f(\beta)- \theta_0\Tau]+\gamma_\cL$.

The annealed entropy can be calculated
from $\AnnealFixedLengthF$, with\footnote{To 
focus on the positional entropy, we did not include here the contribution 
due to loop entropy, i.e., we
treated $\gamma_\cL$ as an energy term despite its entropic origin.}
\begin{equation}
\AnnealFixedLengthS=-k_B^{-1}{\partial\AnnealFixedLengthF\over\partial T}
= \ln N - \beta[f(\beta)-\eps(\beta)] \cL,
\label{AnnealFixedLengthS}
\end{equation}
where 
$\eps(\beta) \equiv - \frac{\partial}{\partial \beta} \ln \Lambda.$
It will also be useful to introduce the 
{\em relative entropy} per base for the fixed-length
bubble,
\begin{equation}
\FixedLengthH(\beta) \equiv [\ln N - \AnnealFixedLengthS]/\cL
= \beta[f(\beta)-\eps(\beta)].
\label{RelEntropy}
\end{equation}
Note that 
being the difference between $f$ and $\eps$, the quantity 
$\FixedLengthH$ is a measure of the {\em intrinsic variation} 
in the binding energies $\dG$'s for a random sequence with nucleotide
frequency $p_b$, and is independent of the average
binding energy $\AvedG$ which external environment such as
the temperature or solvent conditions most directly affect.

Derrida's solution of REM shows that the annealed entropy
$\AnnealFixedLengthS$ vanishes at the transition to the
``low-temperature'' phase, beyond which the annealed approximation 
is no longer applicable. Using Eqs. (\ref{AnnealFixedLengthS})
and (\ref{RelEntropy}), we can write the condition for phase 
transition as
\begin{equation}
\cL_{\rm loc} = \ln N/\FixedLengthH(\beta),
\label{glass.cond}
\end{equation}
which gives the minimal bubble size for localization at a given $N$.
With the values of $\dG$'s obtained from Ref.~\cite{SantaLucia}
%at the physiological condition of $[{\rm Na}^+] = 0.0745~{\tt M}$ 
at  $[{\rm Na}^+] = 1~{\tt M}$ and $37 \Celsius$,
and assuming an equal nucleotide distribution (i.e., $p_b=1/4$ for 
all bases), we find
%$\beta\tf \approx  1.85 - 2.26 \Tau/\Gd$ and
%$\beta\te \approx  1.50 - 2.26 \Tau/\Gd$, making
$f\approx 1.83\, \kT$, $\eps\approx 1.50\,\kT$, so that
$\FixedLengthH \approx 0.33$ and $\cL_{\rm loc} \approx 20\, \bp$
for $N \sim 10^3\,\bp$. 
From Eq.~(\ref{glass.cond}), it is clear that as $N\to \infty$
any fixed-length bubble remains {\em delocalized}.

\subsection{Bubble without Length Constraint}

The full partition function $\SingleBubbleZ$ is obtained simply 
by summing $\FixedLengthZ_L$ for different $L$'s. We will 
again approach the problem by first applying the annealed 
approximation and then determining where it breaks down.\\

\noindent{\bfseries \sffamily \small 1.~Annealed approximation:}  
%\subsubsection{Annealed Approximation:}
The annealed partition function
$\AveSingleBubbleZ(N) \equiv \sum_{L=1}^N \AveFixedLengthZ_L$
has a transition at $\Tau_{\rm a}=f(\beta)/\theta_0$, where
the exponential factor in (\ref{AveFixedLengthZ}) reaches one:
The sum over $L$ is finite and $\AveSingleBubbleZ \propto N$
only if $\Tau<\Tau_{\rm a}$.  
In this regime, the annealed free energy is simply 
$\AnnealSingleBubbleF \approx -\kT \ln N + \init.$
The annealed energy 
$\AnnealSingleBubbleE \equiv -\frac{\partial}{\partial\beta}
\ln\AveSingleBubbleZ(\beta)$ is also readily computed; it can be
expressed as
$\AnnealSingleBubbleE = \init + [\eps(\beta)-\theta_0\Tau]\cdot
\AnnealSingleBubbleL$ where $\AnnealSingleBubbleL(\Tau) \equiv 
\sum_{L=1}^{N\to\infty} L\,
\AveFixedLengthZ_L/\AveSingleBubbleZ$
is the average bubble length in the annealed approximation.
As $\Tau$ approaches $\Tau_{\rm a}$, $\AnnealSingleBubbleL(\Tau)$
diverges, and the annealed entropy
\begin{equation}
\AnnealSingleBubbleS= \ln N -\beta[\theta_0\Tau-\eps(\beta)]\cdot
\AnnealSingleBubbleL
\label{AnnealSingleBubbleS}
\end{equation}
becomes negative.

In the limit $N \to\infty$,
the anealed free energy $\AnnealSingleBubbleF$ is actually identical to
 $\AveSingleBubbleF$ for all $\Tau\leq\Tau_{\rm a}$.
%To show that the two quantities share 
%the same leading order term, let us consider 
This can be seen from the inequalities
$\overline{\ln\FixedLengthZ_{L=1}}\leq\overline{\ln \SingleBubbleZ}
\leq \ln\AveSingleBubbleZ,$
and $\FixedLengthZ_{L=1}>N\min\{
\exp[-\beta \DG_1]\}$. Since both the lower and upper bounds grow as $\ln N$, 
\begin{equation}
\AveSingleBubbleF\equiv -\kT \overline{\ln \SingleBubbleZ}\simeq -\kT \ln N
\label{AvesingleBubbleF}
\end{equation}
for all $\Tau\leq\Tau_{\rm a}$.
\\

\noindent{\bfseries \sffamily \small 2.~Ground-state properties:}
%  
%We next describe the behavior of the bubble in the localized
%phase, i.e., for $\Tau > \Tau_{\rm loc}$. 
%When $\Tau$ exceeds the localization threshold $\Tau_{\rm loc}$,
%the system is dominated by one or a few low energy states
%which correspond to stretches of exceptionally 
%high \AT-content. 
To find the ground-state of the unconstrained bubble
in a long random sequence, we need to study the statistics
of stretches of exceptionally high \AT-content. 
If we neglect the polymeric contribution $\gamma_L$ to the bubble energy
[to be justified shortly], then the ground-state
energy $\MinE$ expected in a sequence of length $N$ 
can be computed exactly from large-deviation 
theory~\cite{KarlinAltschul,Dembo}, with
\begin{equation}
\AveMinE(N) \approx -\lambdainv \ln N.
\label{MinE}
\end{equation}
The constant $\lambda$ in Eq.~(\ref{MinE}) 
can be expressed as the unique positive root of the equation
\begin{equation}
f(\lambda)=\theta_0\Tau,
\label{f-lambda}
\end{equation}
where $f$ is defined by the $\dG$'s through Eqs.~(\ref{M}) and (\ref{f}).
Note that, at $\Tau=\Tau_{\rm a}$, Eq. (\ref{f-lambda}) is satisfied
with $\lambda=\beta$. In this case, (\ref{MinE}) coincides with
(\ref{AvesingleBubbleF}). 

The length of the minimal energy bubble is also known from 
large-deviation theory~\cite{Dembo,YuHwa}, with 
\begin{equation}
\MinL(\Tau) = \ln N/\MinH(\Tau),
\label{glass.cond3}
\end{equation}
where the relative entropy $\MinH$ is given exactly by 
\begin{equation}
\MinH(\Tau)= \lambda(\Tau) \cdot [\theta_0\Tau - \eps(\lambda)].
\label{MinH}
\end{equation}
From the logarithmic dependence of the bubble length $\MinL$ on $N$,
it is clear that the corresponding 
polymeric contribution $\gamma_{\MinL} \sim \ln (\ln N)$ can indeed
be treated as a constant shift of bubble energy.\\

\noindent{\bfseries \sffamily \small 3.~Phase transitions:}  
Based on the above discussion, a phase transition can be formally
established in the limit $N\rightarrow\infty$. This is seen by 
comparing the expressions (\ref{AvesingleBubbleF}) and (\ref{MinE}).
For $\Tau > \Tau_{\rm a}$, solution to (\ref{f-lambda}) satisfies
$\lambda<\beta$. Consequently, (\ref{AvesingleBubbleF}) must
break down there, yielding a phase transition at 
$\Tau=\Tau_{\rm a}$. Since $\AveSingleBubbleF \le \AveMinE$ in general 
(e.g., for all $\Tau > \Tau_{\rm a}$), and at the phase transition point 
$\Tau = \Tau_{\rm a}$ the equality  $\AveSingleBubbleF = \AveMinE$ 
already holds, i.e., the ground-state already dominates, then we must 
have the ground-state dominating throughout the localized phase. This 
is exactly the behavior of the random energy model~\cite{rem}.
 
A physical understanding of the transition can be obtained by
examining the importance of the ground-state contribution 
$\exp(-\beta\MinE)\sim N^{\beta/\lambda}$ to the
partition sum $\SingleBubbleZ$ as the applied twist $\Tau$ is varied.
For $\Tau<\Tau_{\rm a}$, the ratio $\beta/\lambda(\Tau)$ is less than one.
In this case, the energy gain $\AveMinE(N)$ of placing
the bubble at the site of the lowest energy is insufficient to
overcome the entropy $\ln N$ of placing the bubble in different positions,
hence the bubble is typically small and delocalized.
When $\Tau$ exceeds $\Tau_{\rm a}$, the ground-state contribution
grows faster than $N$, signalling dominance of one or a few
low energy states where the bubble typically resides.
The transition is thus identified as the localization transition
%\footnote{One might contemplate a ``weakly-localized'' phase
%of $M$ states contributing significantly to $\SingleBubbleZ$,
%with $1\ll M\ll N$. This is in fact not likely owing to the following
%consideration. Let each such state be the ground-state in a sequence
%of length $N/M$, their total contribution to $\SingleBubbleZ$
%is estimated to grow as $M(N/M)^{\beta/\lambda}$.
%Maximizing with respect to $M$, we obtain $M\sim 1$ for $\lambda<\beta$
%and $M\sim N$ for $\lambda>\beta$. Thus such ``weak-localization'' is
%not expected to occur except possibly at the transition point
%$\Tau=\Tau_{\rm a}$.}
of the bubble at $\Tau_{\rm loc}=\Tau_{\rm a}$.

The onset of the zero entropy point can be obtained
from Eq.~(\ref{AnnealSingleBubbleS}) and written as
\begin{equation}                                                         
\ln N = \SingleBubbleH(\beta,\Tau) 
\cdot \AnnealSingleBubbleL(\beta,\Tau),
\label{glass.cond2}
\end{equation}
where 
\begin{equation}
\SingleBubbleH(\beta,\Tau) = \beta\cdot[\theta_0\Tau-\eps(\beta)]
\label{SingleBubbleH}
\end{equation}
is the relative entropy of the unconstrained bubble.
These equations are analogous to the expressions
(\ref{glass.cond3}) and (\ref{MinH}) for the ground-state
bubble. In fact, both $\MinL(\Tau)$ and $\MinH(\Tau)$
are reproduced through the substitution $\beta\rightarrow\lambda(\Tau)$, 
e.g., $\MinL(\Tau) = \AnnealSingleBubbleL(\lambda(\Tau),\Tau)$.
This turns out to be true also for other thermodynamic variables.
Thus the localized phase at different $\Tau$'s can be viewed
as the phase transition points of systems with different effective
temperatures $\lambda^{-1}(\Tau)$; this will be clearly manifested in the bubble dynamics 
discussed below.

%Unlike the fixed length problem, localization for the unconstrained
%system can occur for arbitrarily large $N$.
%In the limit $N\to\infty$, Eq.~(\ref{glass.cond2}) is satisfied
%with a diverging $\AnnealSingleBubbleL$,  which occurs as
%$f(\beta)-\theta_0\Tau \to 0^+$. 
%This leads to a thermodynamic localization-delocalization transition
%for the unconstrained single bubble system 
%at $\Tau_{\rm loc} = f(\beta)/\theta_0$. For large but finite $N$, 
%the localization condition can be alternatively cast    
%as $\AnnealSingleBubbleL(\Tau_{\rm loc}) = \cL_{\rm loc}$ 
%using Eqs.~(\ref{glass.cond}) -- (\ref{SingleBubbleH}), 
%i.e., {\em  localization occurs for sufficiently large applied torque when
%the size of the  unconstrained
%bubble reaches the critical size $\cL_{\rm loc}$.}

Next we observe that since $\MinH \propto \lambda$ 
[see Eq.~(\ref{MinH})], the bubble length
diverges (or approaches $N$) as $\lambda\to 0$. 
This defines the point of bulk denaturation\footnote{Note however 
that the helical segments separating adjacent bubbles can be
stable even beyond $\Gd$, so that complete separation of the two
strands takes place at $\Tau > \Gd$. } $\Gd$,
i.e., 
\begin{equation}
\theta_0 \Gd \equiv \lim_{\lambda\to 0} f(\lambda)=\AvedG,
\label{Gd-def}
\end{equation}
where the second equality is obtained from manipulating Eqs.~(\ref{M})
and (\ref{f}). Using $\AvedG \approx 1.40\ \kT$ [derived from the $\dG$'s
in Ref.~\cite{SantaLucia}], we find 
$\Gd \approx 10\ {\rm pN}\cdot{\rm nm}$.
The dependence of $\lambda$ on $\Tau$ close to $\Gd$ can be obtained
from the expansion
\begin{equation}
f(\lambda)=\AvedG - \frac{\lambda}{2}\AvedGG + O(\lambda^2).
\label{low-lambda}
\end{equation}
Inverting the above for $\lambda$ using 
(\ref{f-lambda}) and (\ref{Gd-def}), we find
\begin{equation}
\lambda(\Tau) = \frac{2\theta_0}{\beta \AvedGG}\
\big(\Gd-\Tau\big) + O\big(\Gd-\Tau\big)^2.
\label{lambda-approx}
\end{equation}
It turns out that the term linear in $\Gd-\Tau$  in (\ref{lambda-approx})
already gives a very good approximation (to within $1\%$) of $\lambda$ 
throughout the
localized phase where $\lambda/\beta < 1$. The localization transition
point $\Tau_{\rm loc}$ can be thus obtained by solving 
Eq.~(\ref{lambda-approx}) with $\lambda(\Tau_{\rm loc})=\beta$. 
Using $\beta^2\AvedGG \approx 0.565$ [derived from Ref.~\cite{SantaLucia}], 
we find $\Gd-\Tau_{\rm loc} \approx 2\ {\rm pN}\cdot{\rm nm}$.
%
%\begin{equation}
%\Gd-\Tau_{\rm loc} \approx 0.46\ \kT \approx 2\ {\rm pN}\cdot{\rm nm}.
%\label{Tau-loc-num}
%\end{equation}
Unlike the value of $\Gd$ which is derived from the average
stacking energy $\AvedG$ and hence sensitive to  temperature,
ionic strength, etc., the difference $\Gd-\Tau_{\rm loc}$
is set by the variance of $\dG_{b,b'}$ and should be much less
sensitive to experimental conditions. 
The same is expected for the
relative entropy, which has the form 
\begin{equation}
\MinH(\Tau) \approx 2 \theta_0^2 (\Gd-\Tau)^2/\AvedGG
\label{MinH-approx}
\end{equation}
throughout the localized phase.

\subsection{Multiple bubbles}

The localization transition discussed above 
occurs only as $N\to\infty$. However
for large $N$, the single-bubble approximation 
will break down regardless of the large (but finite) 
bubble cost $\gamma_L$. When multiple bubbles are localized, 
each bubble is effectively in a finite-length 
system, thereby blurring the localization transition. 

We first analyze the delocalized phase for which the annealed
approximation is valid. Once multiple bubbles are allowed in 
the system, we expect a broad range of bubble lengths,
as described by the distribution (\ref{AveFixedLengthZ}).
Qualitatively, we expect only the largest bubbles, of size 
$\AnnealSingleBubbleL(\Tau)$ to be localized as 
$\Tau \to \Tau_{\rm loc}$, while the smaller ones remain delocalized.
We shall thus focus on these large bubbles: 
It is the average separation distance 
$\Nx$ between these large bubbles that sets 
the effective system size
of the single-bubble localization problem.

The Boltzmann weight of one such large bubble in a sequence of
length $N\gg \AnnealSingleBubbleL$ is 
$\W(N) = e^{-\beta\init} N/\AnnealSingleBubbleL^{\alpha}$
in the vicinity of the localization transition.
Setting $\W(N)=1$ yields the typical 
spacing between the large bubbles on the delocalized side,
\begin{equation}
\AnnealNx \approx 
e^{\beta\init}\,\AnnealSingleBubbleL^\alpha(\Tau).
\label{Nx.1}
\end{equation}
Note that for bubbles of size $10\,\bp$'s, the crossover length
is already of the order of $10^3\,\bp$'s.
A similar estimate can be made on the localized phase
using the exact expression~\cite{KA93} 
for the lowest energy for multiple bubbles. We find
%
%In the localized phase, the situation is different.
%Suppose we have $k$ localized bubbles, with total energy 
%$\MinE_k(N)$. Using the probability distribution 
%of $\MinE_k$ derived  by  Karlin and Altschul~\cite{KA93}, we find
%\begin{equation}
%\overline{\MinE_k}(N) \approx 
%-\lambdainv\, [k \; \ln N + k (1-\ln k)] + k \gamma_{\MinL}.
%\end{equation}
%The optimal number of bubbles $\Mink$ in the system can be 
%obtained by minimizing $\overline{\MinE_k}$ with respect to $k$, giving
\begin{equation}
\MinNx %= N/\Mink(N) 
\approx 
\bigg[e^{\beta\init}\,(\MinL)^\alpha\bigg]^{\lambda(\Tau)/\beta}.
\label{Nx.2}
\end{equation}
as  the average distance between large bubbles of size $\MinL$.

 For $N\gg \Nx$, 
the system consists of $N/\Nx$ effective number
of single-bubble subsystems, each of length $\Nx$.
At the localization ``transition'' of an infinite system then,
we have $\ln \AnnealNx 
= \SingleBubbleH(\beta,\Tau_{\rm loc}) \AnnealSingleBubbleL(\Tau_{\rm loc})$ 
[see Eq.~(\ref{glass.cond2})]. Together with Eq.~(\ref{Nx.1}) 
[or (\ref{Nx.2}) with $\lambda=\beta$],
we find $\AnnealSingleBubbleL(\Tau_{\rm loc})\approx 25\,\bp$  
at the onset of localization (using 
$\init\approx 3\,\kT$ and $\SingleBubbleH \approx 0.33$), with
the crossover length $\AnnealNx \approx 6500\,\bp$.
Thus we expect there to be typically one bubble of $\sim 25\bp$
in a random DNA double-strand of length $\sim 6500\,\bp$'s
at the localization transtion.

\section{Bubble Dynamics}

The localization of bubbles is reflected ultimately in their
slow dynamics.
%through the above thermodynamic analysis,
We expect bubbles to diffuse
freely along the DNA double-helix in the delocalized phase,
but become trapped in low energy positions
in the localized phase. Details of the bubble movement in the latter case, 
however, can be rather complicated with nontrivial memory (or ``aging'')
effects typical of glassy states~\cite{aging1,aging} as will be described
below.

\subsection{Model}

For simplicity, we will restrict ourselves to the description of
the movement of a single-bubble over its lifetime, which can be rather
long in the localized phase. For reasons discussed above, interaction
with other bubbles can be neglected when the bubble displacement
is within a distance of order $N_\times \sim 10^3\, \bp$. 
We will also neglect the polymeric loop entropy $\gamma_L$ which 
provides essentially a constant shift to the bubble energy
as shown in the single bubble section. 
%Under these assumptions,
%the bubble dynamics is quite similar to that of 
%a particle diffusing in a one-dimensional random potential,
%which has been studied quite extensively in the past
%(e.g., in ~\cite{berbou} and in~\cite{aging}).

In addition to the drift and breathing motion, a bubble may also
shrink to zero size and disappear from the system. To our knowledge,
the time scale involved for the spontaneous collapse of a bubble,
particularly under an applied twist, has not been documented so far.
Zipping the bubble requires not only pairing of the bases in the open
segment, but also rewinding of the helix against the applied undertwist,
both of which contribute to the energy barrier to the no-bubble state.
This suggests a long lifetime for a bubble, which can be enforced by 
setting a lower bound 
(e.g., $10\,\bp$) in the allowed bubble length. However, as we will
see, the long-time behavior of bubble dynamics 
is determined crucially by the occurrence of
the large bubble states, and insensitive to the value of the lower bound
on $L$, as long as the $L=0$ state is excluded.
Once accurate estimates of bubble lifetime become available, one
may supplement the discussion below with such a cutoff.

\subsection{Scaling Theory}

Equation (\ref{MinE}) gives the lowest energy of an unconstrained
bubble in a sequence of length $N$, while a bubble with its 
position (but not size) fixed typically has an energy of the order
$\lambda^{-1}$ for $\Tau<\Tau_{\rm loc}$.
 For small $\lambda$, the energy variation
$\Delta E(N)\simeq \lambda^{-1}\ln N$ is large, hence
the bubble dynamics is dominated 
by the thermal escape from the deepest trap.
The escape time is thus $t_{\rm e}(N) \simeq e^{\beta\Delta E(N)} 
\sim  N^{\beta/\lambda}$, i.e., the dynamics is {\em sub-diffusive} 
deep in the localized phase (where $\beta\gg \lambda$).

To investigate the dynamical behavior in more detail, especially
close to the localization transition where $\lambda \approx \beta$,
we need to include also the random motion of the bubble along the
double strand. Towards this end, it is useful to describe the 
bubble dynamics as a {\em single} point moving 
in the two-dimensional space 
spanned by the bubble's only two degrees of freedom,
its instantaneous length $L$ and the position of one of its ends, say $m$. 
The statistics of the two-dimensional
energy landscape $\DG_L(m)$ is well-characterized by the large-deviation
theory~\cite{YuHwa}. It  consists of a number of valleys, whose depths 
(denoted by $\ValleyG$'s) are given by the  Poisson distribution
\begin{equation}
{\cal D}\left(\left|\ValleyG\right|\right) \propto e^{-\lambda\left|\ValleyG\right|},
\label{barrier-dist}
\end{equation}
where $\lambda$ is the constant defined through (\ref{f-lambda}). The
typical valley length is $\ValleyL \sim 1/\MinH$, where $\MinH$
is given by (\ref{MinH}).
The valleys are spread out along the corridor at $L \lesssim \ValleyL$, 
separated by a typical distance $M$ which is also calculable from
the large-deviation theory. For much larger $L$'s, the bubble energy
becomes prohibitively high.

Clearly, the dynamics consists of two parts: At short times,
it is dominated by the escape of the
bubble out of an individual valley, and is analogous to the
(biased) Sinai problem~\cite{Sinai}. At longer time scales,
the bubble ``hops'' from one valley to another along the corridor
of valleys. 
This dynamics, which is essentially that of a particle
traversing a series of exponentially-distributed energy valleys 
[see Eq.~(\ref{barrier-dist})],
has been extensively investigated previously in the context
of the one-dimensional ``trap'' model~\cite{slowa,slowb}.
Here we review some key results  and refer the readers to 
Ref.~\cite{berbou} for details.
The basic quantity is the time
%\begin{equation}
$\tau(\ValleyG) \propto e^{\beta\left|\ValleyG\right|}$
%\label{tau}
%\end{equation}
to escape each valley of depth $\ValleyG$. 
The average time to traverse $K$ valleys
over a length scale $N=K\cdot M$
by random walk is then given by 
\begin{equation}
t_{\rm e} = K^2 \cdot \langle \tau \rangle 
\sim N^2 \int_{\lambdainv}^{|\AveMinE|} 
dx \, \tau(x)\, {\cal D}(x),
\label{T-def}
\end{equation}
where $\langle \tau \rangle$ is the average of the trap time 
$\tau(\ValleyG)$, 
and the limits of integration in (\ref{T-def}) are from 
the magnitude of the typical valley
depth $\lambdainv$ to that of the deepest valley (\ref{MinE})
expected for a segment
of length $N$.

The total time according to Eq.~(\ref{T-def}) can be written as
$t_{\rm e}(N) \propto N^z$, with the dynamic exponent $z$ given by
\begin{equation}
z = \left\{
\begin{tabular}{ll}
$2$ & \quad for ~ $\lambda > \beta$ ~ (or $\Tau<\Tau_{\rm loc}$)\\
$1+\beta/\lambda$ & \quad for ~ $\lambda < \beta$ ~ (or $\Tau_{\rm loc} 
< \Tau < \Tau_{\rm d}$)
\end{tabular}\right.
\label{z-exp}
\end{equation}
The anomalous exponent $z>2$ in the glass phase 
shows explicitly that the dynamics is slow, i.e., 
{\em sub-diffusive}. 
% A well-known glassy system exhibiting
%sub-diffusive dynamics is the 1+1 dimensional vortex glass~\cite{vg}.  
%It is clear from the above analysis that the origin of dynamic
%anomaly lies in the logarithmically diverging barriers as had been
%suspected for the vortex glass.

\subsection{Glassy Dynamics}

We next report the result of a Monte-Carlo simulation of the bubble
dynamics on predefined random nucleotide sequences. We impose {\em
local dynamics} in which the bubble can only change its length $L$ or
shift its end position $m$ by a single base,
% except if the move is not
%consistent with keeping a positive length (i.e. 
as long as $L\ge 1$. To remove
edge effects and probe the asymptotic dynamics, we use a very large
sequence length ($> 10^4\ \bp$) so that the bubble never reaches the
boundary of the sequence given the duration of our numerical study.
%After proposing a legitimate move, we determine whether or not to
%accept it by using the standard Metropolis algorithm: Let the change
%in bubble energy $\DG_L(m)$ due to a proposed move be $\delta E$.  The
%move is always accepted if $\delta E \le 0$, and is accepted with
%probability $e^{-\beta \delta E}$ if $\delta E > 0$.  
%This procedure
%is expected to approach the equilibrium distribution $e^{-\beta
%\DG_L(m)}$ in the long time limit (although in practice, reaching
%equilibrium may take a very long time in the glassy regime.)  In this
%way, we analyzed our system throughout the localized phase, i.e., for
%$\Tau_{\rm loc} < \Tau < \Gd$.  
All disorder-averaged quantities reported are
performed over $10^4$ random sequences. 
% This kind of slow decays is
%discussed in detail in~\cite{berbou}. We have checked that different
%detailed local rules lead to the same asymptotic behavior. 
%Among others we have observed that if we allow the bubble to disappear
%when it reaches zero length and to reappear in a random position on
%the chain the asymptotic properties that we discuss here do not
%change: they are determined from a regime where the bubble is already
%large, and there this kind of effects do not play a role.
\\

\begin{figure}[t]

\centerline{\includegraphics[width=3.1in, angle=0]{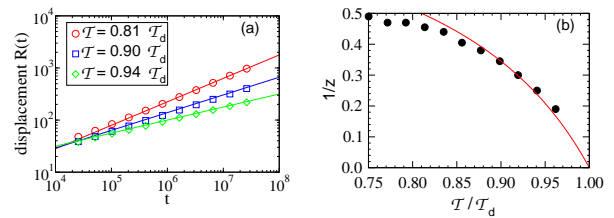}}
  \caption{(a) Average bubble position vs time for various values
of $\Tau$'s in the glass phase: the solid lines are power law fits. 
(b) The extracted exponents vs $\Tau$: the solid line is
the prediction of the scaling theory Eq.~(\ref{z-exp}).
    \protect\label{F-COM}
  }
\end{figure}

\noindent{\bfseries \sffamily \small 1.~Anomalous diffusion:}  
To characterize the slow dynamics quantitatively, we show in
Fig.~\ref{F-COM}(a) the time evolution of the average displacement 
$R(t) = |m(t)-m(0)|$
of the bubble position for a few selected values of $\Tau$'s 
in the glass phase.
Clearly, the displacement can be described by a power law of the
form $R(t)\propto t^\nu$, where we expect $\nu = 1/z$.
In Fig.~\ref{F-COM}(b), we plot the extracted exponents (circles) 
for different values of $\Tau$'s in the range 
$\Tau_{\rm loc} \le \Tau < \Gd$. The expected
values $1/z$ according to Eq.~(\ref{z-exp})
(using the linear expression in (\ref{lambda-approx}) for $\lambda$)
is shown as the solid line for comparison.
We note that the observed exponents follow the general trend
predicted, changing continuously from $1/z = 0.5$ close to
the expected location of the glass transition ($\Tau_{\rm loc}\approx 0.8 
\Gd$), towards zero as $\Tau \to \Gd$. 
For $\Tau$ close to $\Gd$, the dynamics becomes exceedingly slow, 
making it difficult to access the asymptotic region. 
For $\Tau \approx \Tau_{\rm loc}$, we also observed some 
finite-size effect. The overall agreement between the scaling theory and
numerical results is within $5\sim 10\%$ over the range tested.

\begin{figure}[b]
\centerline{\includegraphics[width=3.1in, angle=0]{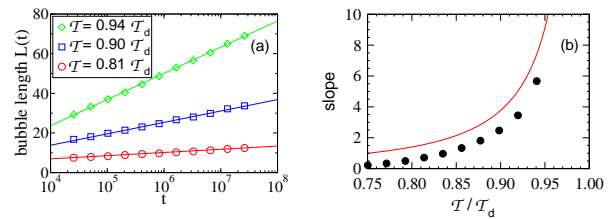}}
  \caption{(a) Average bubble length vs time for several $\Tau$'s; 
the solid lines are fits to the form $L(t) = a + b\, \ln t$.
(b) Slope $b$ of the logarithmic time-dependence of $L(t)$ for
various $\Tau$'s.  The solid line is the corresponding quantity
for the upper bound of the bubble length $\MinL(t)$; see text.
    \protect\label{F-LENGTH}
  }
\end{figure}

In Fig.~\ref{F-LENGTH}(a), we show the dependence of the 
average bubble length on time for different $\Tau$'s.
The data depict the slow, logarithmic growth of the bubble length. 
Logarithmic growth 
is one of the signatures of glassy dynamics. Its occurrence
in this particular system can be understood quantitatively 
as follows: The optimal bubble size $\MinL(N)$ in a segment of
length $N$ depends logarithmically on $N$; see Eq.~(\ref{glass.cond3}). 
On the other hand, for a bubble placed at an arbitrary position
in a long sequence, the {\em effective} sequence length 
is the distance the bubble can explore within a time $t$, 
i.e., $N \sim t^{1/z}$ for the sub-diffusive dynamics expected
in the glassy regime. Hence, 
\begin{equation}
\MinL(t) \approx \frac{1}{z\cdot\MinH} \ln t + {\rm const.}
\label{L-vs-T}
\end{equation}
is the expected length of the optimal bubble within a time t.
Generally, we expect  $\MinL(t)$ to be the upper bound of the
observed bubble length $L(t)$, with $L(t) \approx  \MinL(t)$
for large $t$ deep in the glass phase. 
However below the glass transition,
$L(t)$ must be finite even for $t\to\infty$.

In Fig.~\ref{F-LENGTH}(b), we show the coefficients of the
observed logarithmic time dependence of $L(t)$ for $\Tau$'s
throughout the range $\Tau_{\rm loc} < \Tau < \Gd$. Also shown
is the upper bound $1/(z\cdot \MinH)$ (solid line) 
according to (\ref{L-vs-T}), using the expression 
(\ref{MinH-approx}) for $\MinH$.
We note that the difference
between the data (circles) and the upper bound is nearly constant
($\approx 1$) for the range studied.\\

\noindent{\bfseries \sffamily \small 2.~Aging:}  
Perhaps the most characteristic feature of
glassy dynamics is that the system ``ages'', e.g., 
the temporal fluctuation of the system depends
on how long the system has evolved from some (arbitrary)
initial condition~\cite{aging1,aging}: The longer it has evolved, the 
slower it
fluctuates. This is easy to understand in the context of a rough
energy landscape with deep valleys and high barriers, since
the longer the system evolves, the deeper the energy valley it
finds, and hence the higher the barrier it will have to overcome
to travel farther. This feature is in marked contrast to sub-diffusive
hydrodynamic systems which are time-translationally invariant.

Quantitatively, we can define the aging phenomenon via the
time-dependent correlation function $C(t_{\rm w},\dt)$, which
measures how much the system changes in time $\dt$,
after first evolving for  a {\em waiting} period $t_{\rm w}$ 
from the initial condition. Let us
define a binary variable $\eta_i(t)\in \{0, 1\}$, for each 
base $i$ of the nucleotide sequence. 
$\eta_i(t)$ takes on the value $1$ 
if  base $i$ is open and belongs to the bubble at time $t$, 
and the value $0$ if base $i$ is paired.
The correlation function, defined as
%\begin{equation}
$C(t_{\rm w},\dt)\equiv  \sum_i  \eta_i(t_{\rm w})\eta_i(t_{\rm w}+\dt)$
%\end{equation}
after averaging over $10000$ random sequences, is a measure
of the average {\em self-overlap} of the bubble 
at time $t_{\rm w}$ and $t_{\rm w}+\dt$.
A more convenient quantity to characterize is the fraction
of overlap, $C(t_{\rm w},\dt)/L(t_{\rm w})$, 
where $L(t)= \sum_i \eta_i(t)$ is the instantaneous bubble length.

\begin{figure}[b]

\centerline{\includegraphics[width=3.1in, angle=0]{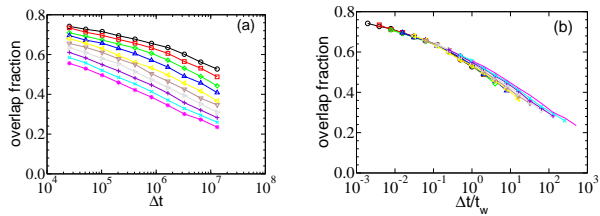}}
 \caption{ Aging plot: (a) Average overlap fraction $C(t_{\rm w},\dt)/L(t_{\rm w})$
for different $t_{\rm w}$'s (from $1\times$ to $512\times$ $25,600$ Monte-Carlo
steps) deep in the glass phase, with
 $\Tau = 0.9\Gd > \Tau_{\rm loc}$;  (b) Scaling plot of (a) with $\dt$
normalized by $t_{\rm w}$.
    \protect\label{F-AGING}
  }
\end{figure}

In Fig.~\ref{F-AGING}(a), we show the overlap fraction,
parameterized by the different waiting time $t_{\rm w}$'s for the system
biased deep in the glass phase with $\Tau = 0.9\ \Gd$.
 The overlap fraction clearly depends
on the waiting time, illustrating the glassy nature of the dynamics.
In contrast, the same quantity computed for $\Tau < \Gd$ 
(data not shown) gives no statistically significant dependence 
on $t_{\rm w}$. To characterize more quantitatively the behavior, 
we re-plot in Fig.~\ref{F-AGING}(b) the curves in (a) with
$\dt$ normalized by $t_{\rm w}$. We find these curves to collapse
reasonably onto a single master curve which exhibits 
a weak kink at $\dt/t_{\rm w} \sim 1$.
 A naive explanation of this behavior is
that for $\dt \ll t_{\rm w}$, the bubble stays approximately within 
the energy valley found at time $t_{\rm w}$,
while for $\dt \gg t_{\rm w}$, the bubble makes excursion far away
from the valley. For the one-dimensional ``trap'' model, 
it was shown rigorously~\cite{corr-scale} that
$C(t_{\rm w}, \dt)$ indeed scales as a function of $\dt/t_{\rm w}$, 
even though the largest trap time actually scales sub-linearly with
$t_{\rm w}$. This behavior can be understood in terms of the particle
making multiple returns to the original valley after escaping 
it~\cite{berbou}, as manifested by the slow decay shown in 
Fig.~\ref{F-AGING}(b) for $\dt \gg t_{\rm w}$.

\section{Discussion}

In this study we investigated the thermodynamic and dynamic 
behaviors of twist-induced denaturation bubbles in a long, 
random sequence of DNA. The small bubbles associated with weak twist
are delocalized, e.g., they flicker in and out of existence according
to the Boltzmann distribution and are independent of the DNA sequence.
The bubbles increase in lengths
upon increase in the applied torque. When the largest
bubbles reach a critical size $\cL_{\rm loc}$ which is of the
order of a few tens of bases, the bubbles become localized to
\AT-rich segments which occur statistically in a long random sequence.
According to the parameters~\cite{SantaLucia} taken at
$37\Celsius$ with  $[{\rm Na}^+] = 1~{\tt M}$, the localization 
``transition'' occurs at 
$\Tau_{\rm loc} \approx 8 \ {\rm pN}\cdot{\rm nm}$,
which is $\sim 80\%$ of the torque needed for bulk denaturation $\Gd$.
In the localized regime, the bubbles exhibit ``aging'' and
move along the double helix sub-diffusively, with continuously varying 
dynamic exponents.

All of the results are obtained under the single-bubble approximation.
Thermodynamically, we expect this approximation to be valid for
DNA sequences of several thousand bases or less. This is due to
the strongly cooperative nature of bubble formation, as manifested
in the large initiation energy $\init$. The single bubble 
description of dynamics is further restricted by the finite
life time of the bubble: Even at length scales where the single-bubble
approximation is appropriate thermodynamically, the bubble may
annihilate and reappear elsewhere in the sequence, effectively
performing long-distance hops. %~\cite{Frank-Kamenetskii}. 
Experimental knowledge of the bubble life time in the presence of 
an applied twist
is needed to estimate the crossover time to the long-distance
hopping regime. Qualitatively, we expect these bubbles to have much
longer life times than the thermally denatured bubbles, since
the applied twist 
plays the role of an energy barrier preventing bubble annihilation.

Finally, we note that  bubble localization characterized in 
this study is a reflection of the statistical background
present in long random nucleotide sequences. This background
 traps the bubble kinetically if the bubble size becomes
sufficiently large. Thus, to localize denaturation bubbles at 
appropriate locations specified by designed sequences (e.g., 
promoters or replication origins)  for biological functions, 
it is necessary
to operate away from the localized regime, i.e., below the 
onset of localization.

\vspace{12pt}

%\section*{Acknowledgement}

This collaboration was made possible by the program on ``Statistical
Physics and Biological Information'' hosted by the Institute for 
Theoretical Physics in Santa Barbara.
The authors benefitted from discussions with D.\ Bensimon,
R.\ Bundschuh, H.\ Chate, U.\ Gerland, D.\ Lubensky, M.\ Mezard, 
and Y.-k.\ Yu.
TH is supported by NSF Grant No.\ 0211308 and a Burroughs-Wellcome
functional genomics award. LT acknowledges the hospitality of 
UC San Diego where part of this work was carried out.
\\

\end{document}